\documentclass{wscpaperproc}
\usepackage{latexsym}
\usepackage{graphicx}
\usepackage{mathptmx}

\usepackage{subcaption}
\usepackage[pdftex,colorlinks=true,urlcolor=blue,citecolor=black,
anchorcolor=black,linkcolor=black]{hyperref}
\usepackage{amsmath}
\usepackage{amsfonts}
\usepackage{amssymb}
\usepackage{amsbsy}
\usepackage{amsthm}
\usepackage[pdftex,colorlinks=true,urlcolor=blue,citecolor=black,anchorcolor=black,linkcolor=black]{hyperref}

\newtheoremstyle{wsc}% hnamei
{3pt}% hSpace abovei
{3pt}% hSpace belowi
{}% hBody fonti
{}% hIndent amounti1
{\bf}% hTheorem head fontbf
{}% hPunctuation after theorem headi
{.5em}% hSpace after theorem headi2
{}% hTheorem head spec (can be left empty, meaning `normal')i

\theoremstyle{wsc}

\begin{document}

%***************************************************************************
% AUTHOR: AUTHOR NAMES GO HERE
% FORMAT AUTHORS NAMES Like: Author1, Author2 and Author3 (last names)
%
%		You need to change the author listing below!
%               Please list ALL authors using last name only, separate by a comma except
%               for the last author, separate with "and"
%

% setting up general page style
\pagestyle{fancyplain}

% setting up page style of first page
\thispagestyle{plain}
\firstPageHead{}

% setting up running header (authors) of subsequent pages
\chead{\fancyplain{}{\itshape Huang, Lam and Zhao}}

% setting up seperation parameters
%\headsep=72pt
\rhead{}
\cfoot{}
\renewcommand{\headrulewidth}{0pt} % (renewcommand needed in fancyhdr to remove top decorative line)
%\headrulewidth=0pt  % ("setlength" needed in fancyheading to remove top decorative line)

%%%%%%%%%%%%%%%%%%%%%%%%%%%%%%%%%%%%%%%%%%%%%%%%%%%%%%%%%%%%%%%%%%%%%%%%%%%%%%
%                                                                            %
%     THESE COMMANDS ARE REQUIRED TO WORK WITH WSC.BST TO MAKE BIBLIO     %
%                                                                            %
%%%%%%%%%%%%%%%%%%%%%%%%%%%%%%%%%%%%%%%%%%%%%%%%%%%%%%%%%%%%%%%%%%%%%%%%%%%%%%
\makeatletter
\let\@internalcite\cite
\def\cite{\def\@citeseppen{-1000}%
    \def\@cite##1##2{(##1\if@tempswa , ##2\fi)}%
    \def\citeauthoryear##1##2##3{##1 ##3}\@internalcite}
\def\citeNP{\def\@citeseppen{-1000}%
    \def\@cite##1##2{##1\if@tempswa , ##2\fi}%
    \def\citeauthoryear##1##2##3{##1 ##3}\@internalcite}
\def\citeN{\def\@citeseppen{-1000}%
%  Pierre L'Ecuyer's fix for multiple cite bug
%  Added by Paul J Sanchez on 4 October 2001
%   \def\@cite##1##2{##1\if@tempswa , ##2)\else{)}\fi}%
%   \def\citeauthoryear##1##2##3{##1 (##3}\@citedata}
    \def\@cite##1##2{##1\if@tempswa, ##2)\else{}\fi}%
    \def\citeauthoryear##1##2##3{##1 (##3)}\@citedata}
\def\citeA{\def\@citeseppen{-1000}%
    \def\@cite##1##2{(##1\if@tempswa , ##2\fi)}%
    \def\citeauthoryear##1##2##3{##1}\@internalcite}
\def\citeANP{\def\@citeseppen{-1000}%
    \def\@cite##1##2{##1\if@tempswa , ##2\fi}%
    \def\citeauthoryear##1##2##3{##1}\@internalcite}
\def\shortcite{\def\@citeseppen{-1000}%
    \def\@cite##1##2{(##1\if@tempswa , ##2\fi)}%
    \def\citeauthoryear##1##2##3{##2 ##3}\@internalcite}
\def\shortciteNP{\def\@citeseppen{-1000}%
    \def\@cite##1##2{##1\if@tempswa , ##2\fi}%
    \def\citeauthoryear##1##2##3{##2 ##3}\@internalcite}
\def\shortciteN{\def\@citeseppen{-1000}%
%  Pierre L'Ecuyer's fix for multiple cite bug
%  Added by Paul J Sanchez on 2 September 2002
%  should have caught this last year...
%   \def\@cite##1##2{##1\if@tempswa , ##2)\else{)}\fi}%
%   \def\citeauthoryear##1##2##3{##2 (##3}\@citedata}
% Shane G. Henderson fix for extra right bracket at end of optional material June 8, 2005
%    \def\@cite##1##2{##1\if@tempswa, ##2)\else{}\fi}%
    \def\@cite##1##2{##1\if@tempswa, ##2\else{}\fi}%
    \def\citeauthoryear##1##2##3{##2 (##3)}\@citedata}
\def\shortciteA{\def\@citeseppen{-1000}%
    \def\@cite##1##2{(##1\if@tempswa , ##2\fi)}%
    \def\citeauthoryear##1##2##3{##2}\@internalcite}
\def\shortciteANP{\def\@citeseppen{-1000}%
    \def\@cite##1##2{##1\if@tempswa , ##2\fi}%
    \def\citeauthoryear##1##2##3{##2}\@internalcite}
\def\citeyear{\def\@citeseppen{-1000}%
    \def\@cite##1##2{(##1\if@tempswa , ##2\fi)}%
    \def\citeauthoryear##1##2##3{##3}\@citedata}
\def\citeyearNP{\def\@citeseppen{-1000}%
    \def\@cite##1##2{##1\if@tempswa , ##2\fi}%
    \def\citeauthoryear##1##2##3{##3}\@citedata}
%
% \@citedata and \@citedatax:
%
% Place commas in-between citations in the same \citeyear, \citeyearNP,
% \citeN, or \shortciteN command.
% Use something like \citeN{ref1,ref2,ref3} and \citeN{ref4} for a list.
%
\def\@citedata{%
    \@ifnextchar [{\@tempswatrue\@citedatax}%
                  {\@tempswafalse\@citedatax[]}%
}

\def\@citedatax[#1]#2{%
\if@filesw\immediate\write\@auxout{\string\citation{#2}}\fi%
  \def\@citea{}\@cite{\@for\@citeb:=#2\do%
    {\@citea\def\@citea{, }\@ifundefined% by Young
       {b@\@citeb}{{\bf ?}%
       \@warning{Citation `\@citeb' on page \thepage \space undefined}}%
{\csname b@\@citeb\endcsname}}}{#1}}%

% don't box citations, separate with ; and a space
% also, make the penalty between citations negative: a good place to break.
%
\def\@citex[#1]#2{%
\if@filesw\immediate\write\@auxout{\string\citation{#2}}\fi%
  \def\@citea{}\@cite{\@for\@citeb:=#2\do%
    {\@citea\def\@citea{, }\@ifundefined% by Young
       {b@\@citeb}{{\bf ?}%
       \@warning{Citation `\@citeb' on page \thepage \space undefined}}%
{\csname b@\@citeb\endcsname}}}{#1}}%

% (from apalike.sty)
% No labels in the bibliography.
%
\def\@biblabel#1{}
\makeatother

%\newlength{\bibhang}
%\setlength{\bibhang}{2em}

% Indent second and subsequent lines of bibliographic entries. Taken
% from openbib.sty: \newblock is set to {}.
% \renewcommand{\refname}{REFERENCES}

\newdimen\bibindent
\bibindent=0.0em
% SEC: was \def\thebibliography#1{\section*{\refname\@mkboth
% SEC: was   {\uppercase{\refname}}{\uppercase{\refname}}}\list
\def\thebibliography#1{\section*{\refname}\list
   {}{\settowidth\labelwidth{[#1]}
   \leftmargin\parindent
   \itemindent -\parindent
   \listparindent \itemindent
   \itemsep 0pt
   \parsep 0pt}
   \def\newblock{}
   \sloppy
   \sfcode`\.=1000\relax}

           % Set up BiBTeX macros

% needed to make the tex document look more like the word counterpart :-(
\setlength{\baselineskip}{12.7pt}

% AUTHOR: Enter the title, all letters in upper case
\title{SEQUENTIAL EXPERIMENTATION TO EFFICIENTLY TEST AUTOMATED VEHICLES}
% SEQUENTIAL EXPERIMENTATION TO EVALUATE AUTOMATED VEHICLES

% AUTHOR: Enter the authors of the article, see end of the example document for further examples
\author{Zhiyuan Huang\\ Henry Lam\\ [12pt]
Department of Industrial and Operations Engineering \\
University of Michigan\\
1205 Beal Avenue\\
Ann Arbor, MI 48105, USA\\
% Multiple authors are entered as follows.
% You may also need to adjust the titlevbox size in the preamble - search for titlevboxsize
\and
Ding Zhao\\[12pt]
Department of Mechanical Engineering\\
University of Michigan\\
2901 Baxter Road\\
Ann Arbor, MI 48109, USA
}

\maketitle

\section*{ABSTRACT}
Automated vehicles have been under heavy developments in major auto and tech companies and are expected to release into market in the foreseeable future. However, the road safety of these vehicles remains a concern. One approach to evaluate their safety is via on-track experimentation, but this requires gigantic costs and time investments. This paper discusses a sequential learning approach based on kriging models to reduce the experimental runs and economize on-track experimentation. The approach relies on a heuristic simulation-based gradient descent procedure to search for the best next test scenario. We demonstrate our approach with some numerical test cases.

\section{INTRODUCTION}                                                                          
\subsection{Background of Automated Vehicles Evaluation}
While automated vehicles (AVs) are currently under intense developments by almost all major auto companies and tech giants, their safety has remained a concern, as reinforced by recent Tesla accidents involving self-driving systems \cite{teslanews}. The difficulty in evaluating AVs is that these vehicles are ``smart", in that they interact with their environments and prompt autonomous actions, and hence cannot be tested using existing standard approaches. 

For example, the so-called test matrix approach, adopted commonly in many vehicle testing procedures, uses fixed and predefined test scenarios to evaluate vehicles. However, an AV producer can tune the algorithm to excel in such test scenarios but fail on others, making the results of the test matrix invalid in capturing the actual risk \cite{Aust2012EvaluationSystems}. In the United States, there are currently no standards or protocols to test AVs with high degrees of automation (known as automation level 2 \cite{national2013preliminary} or higher). Most prospective AV manufacturers at present rely on Naturalist Field Operational Tests (N-FOT) \cite{festa2008festa} to evaluate AV safety, which means putting the vehicle prototypes on actual roads and collecting data from potential accidents or conflicts. Such tests, however, are both time-consuming and costly, as accidents are rare events that can only be assessed under statistical confidence with astronomical road miles driven by these prototypes. According to \shortciteN{akamatsu2013automotive}, an N-FOT ``cannot be conducted with less than \$10,000,000''.

As an alternative, researchers have explored the use of Monte Carlo simulation techniques. \citeN{yang2010development} and \citeN{lee2004longitudinal} evaluated collision avoidance systems by reusing existing N-FOT data, and \shortciteN{woodrooffe2014performance} used forward collision scenarios to evaluate collision warning and mitigation braking technologies on heavy trucks. \shortciteN{zhao2017accelerated}, \shortciteN{huang2017accelerated}, and \shortciteN{huang1001evaluation} applied importance sampling methods to evaluate car-following and lane change scenarios. However,  Monte-Carlo-based methods need to make assumptions on the control and dynamics of AVs. The lack of full knowledge in specifying these assumptions, complicated by the autonomous operations of AVs that are not publicly disclosed, remains one of the key difficulties in carrying out reliable Monte Carlo evaluation. On-track experiments to learn the behaviors of AVs is therefore a crucial step \cite{peng2012evaluation}. These behaviors, once accurately informed, can be used as inputs to the Monte Carlo evaluation. However, such experiments are only recently feasible \cite{mcity} and require huge cost and time investments. This motivates us to explore an adaptive approach to reduce the number of on-track experimental runs needed for the learning. 

% Since the exposure to safety critical scenarios is generally low, Monte Carlo simulations are time-consuming; therefore, variance reduction techniques have been considered. 

% Although variance reduction techniques largely reduced the required number of tests to a scale of thousand, the number is still not affordable for real world on-track experiments. The National Highway Traffic Safety Administration has the responsibility to guarantee the safety of vehicles in the market. On-track experiments with new vehicles are required for safety evaluation (Peng et al. 2012). For Automated Vehicles, the vast number of scenarios in the daily driving make it hard to evaluate. Considering the number of new types entering the market each year, the existing methods are inefficient for the tasks. 

\subsection{Outline of the Sequential Experimentation Approach}

The number of possible scenarios that an AV can react on, which collectively define the behavior of the AV, are typically infinite. This motivates us to consider a metamodel to make our learning feasible. Specifically, we use a kriging framework to model the unknown behavior of AVs, and investigate a myopic approach to sequentially select the next test scenario that can maximize the information gain (thereby reducing the runs needed to achieve a reasonable estimation accuracy). As the gain is in terms of the correctness of the Monte Carlo evaluation, finding the next test scenario generally requires simulation-based optimization. In particular, we investigate a heuristic use of stochastic gradient descent. The simulation is also used to make the final safety evaluation of the AV being considered.

% To overcome the number and cost of experimentation, this paper proposes a Kriging-based sampling approach. The proposed sampling procedure is based on an optimization problem that gives the sample point with the maximum impact to the event probability estimation. We validate the proposed model with numerical simulations and conclude from the results that the approach is an effective replacement for N-FOT and other burdensome evaluative techniques.

Our framework follows from the kriging technique originated from geology (e.g., \citeNP{chiles2009geostatistics}) and further developed in computer experiments (e.g., \shortciteNP{sacks1989design}). The primary use of kriging is to assimilate spatial data under correlation among different design points that is made computationally convenient through Gaussian process modeling. Our approach follows this framework by viewing the test scenarios as design points. In the static settings, the design points are typically selected using space-filling design (e.g., with Latin Hypercube Sampling; \citeNP{kleijnen2008design}). To reduce experimental costs with respect to a specified goal, one can sequentially select the design points, which is the approach we adopt. In particular, we follow the sequential sampling idea that has been applied to sensitivity analysis and optimization \cite{kleijnen2004application,kleijnen2009kriging}. Our work most closely follows the concept of knowledge gradient (e.g., \citeNP{powell2012optimal}) in the Bayesian setting. Other related literature includes the stream of study in stochastic kriging \shortcite{ankenman2010stochastic,staum2009better}, a generalization of the kriging technique to stochastic computer experiments. In this paper, however, we assume the on-track experimentation is error-free and hence relates more closely to the deterministic experimentation framework. On the other hand, stochasticity comes in the evaluation criterion and as a result, as discussed above, our sequential design point search will allude to the use of simulation optimization. 
% Finding the next design point according to this stochastic evaluation gives rise to a simulation optimization problem, for which we adopt an iterative gradient-based method via a heuristic use of stochastic approximation (

% The standard sampling methods are space-filling designs. The most popular method of this type is the Latin Hypercube Sampling (LHS) (Kleijnen 2008). The space-filling designs does not consider the goal of the experiments and hence is considered to be less efficient. Sequential sampling methods are developed for different goals of the experiments. studied sequential sampling methods for sensitivity analysis and  studied the sequential sampling methods for optimization. These approaches are efficient for the specific goals of the Kriging model. Huang et al. (2006a, 2006b) uses optimization in the sampling procedure. This approach used sequential sampling to maximize an augmented expected improvement function, which is connected with the evaluation costs. However, the optimization is not related to a specific use of the model and tries to balance the local and global search for design points. This paper proposes an sampling approach based on optimization which sequentially searches for samples with largest impacts on the event probability estimation. 
% Section \ref{sec:sampling_review} briefly reviews the range of Kriging sampling techniques.
The remainder of this paper is as follows.  Section \ref{sec:example} describes the basic setups in AV evaluation and casts our evaluation framework in the kriging setting. Section \ref{sec:optim} presents our optimization procedure to select test scenarios. Section \ref{sec:numerical} shows some numerical examples. 

% Section \ref{sec:conclusion} concludes this paper.

% \section{Review of Sampling Methods for Kriging} \label{sec:sampling_review}

\section{A KRIGING FRAMEWORK FOR AV EVALUATION} \label{sec:example}
We introduce our framework in two components. First, Section \ref{sec:task} describes the setting and the challenge of AV evaluation and gives a simple illustrative example. Section \ref{sec:kriging} then describes how we cast the AV evaluation task into a kriging-based learning model.

\subsection{The Task of AV Evaluation}\label{sec:task}
\begin{figure}[ht]
							\centering
							\includegraphics[width=0.5\linewidth]{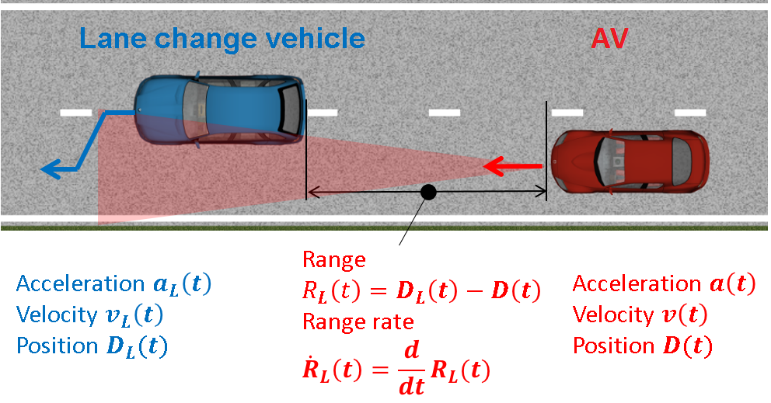}
                            \caption{A lane change scenario.}
                            \label{fig:lane}
\end{figure}

Evaluation of the road safety of AVs requires studying the risk arising from its interaction with the surrounding environments, such as other vehicles driven by human drivers, pedestrians etc. The risk can be measured by probabilistic quantities such as the chance of accidents (e.g., crashes) and conflicts (e.g., the AV and a front car within a dangerously short distance). For example, \shortciteN{zhao2017accelerated} demonstrate this calculation via Monte Carlo simulation with a lane change scenario. Figure \ref{fig:lane} describes this setting, where a human-controlled vehicle driving in front of the AV is cutting into the AV's lane. The AV has a built-in intelligent control system that is assumed deterministic, while the frontal vehicle is susceptible to noisy human behavior and hence is stochastic. 

A collision can occur when the gap is too short at any point of time. Consider a fixed period of time $T$ that represents the typical car-following duration. Denote $R_L(t,\omega)$ as the range between the AV and the human-driving vehicle and $\dot{R}_L(t,\omega)$ as its rate of change, which depend on the physical measurements of both vehicles including accelerations, velocities and positions. $\omega$ denotes the initial condition of the lane change scenario. We say a collision happens if the range at any point of time is too short, say within a threshold $b$. Then the collision probability is $P(R_L(t,\omega)<b\text{\ for some\ }t\in[0,T])$, or equivalently $P(\max_{t\in[0,T]}1/R_L(t,\omega)>1/b)$. 
% Since the lane change is initialized by the frontal human-controlled vehicle, we assume that the condition for the two vehicles are stochastic when the lane change starts. , where $R_L(t)$ denotes the gap between the human-driving vehicle and the AV, and $\gamma$. This quantity depends on the dynamic physical measurements We use the database to fit these three variables to probability distributions.

In general, the stochasticity of the human-driving vehicle, described by its acceleration etc., can be estimated from existing data. \shortciteN{zhao2017accelerated} for instance uses the naturalistic driving data among all the lane change scenarios extracted from the Safety Pilot Model Deployment (SPMD) database \cite{bezzina2014safety}. However, the AV control is typically not fully known to the tester. It could be known by the company that owns its production, but  due to commercial concern such knowledge is not revealed to governmental or public entities who conduct safety tests. So to carry out the Monte Carlo safety test, a governmental unit needs to learn the control system by carrying out its own on-track experiment. This experiment runs on a physical proving ground (e.g., \citeNP{mcity}) which, in the considered setting, can preset the configuration of the frontal vehicle to resemble an actual road condition. Observing how the AV reacts in these conditions provides some information on its underlying intelligent control.
% Fig. \ref{fig:lane} illustrates the scenario and notations of data. 

Other scenarios can be evaluated similarly as above; see, e.g., \shortciteN{zhao2017accelerated} for a car-following setting. In the subsequent discussion we will focus on the lane change situation for illustration. 

\subsection{A Kriging Model}\label{sec:kriging}
% First developed by geostatisticians and further studied by mathematicians, the Kriging model is a powerful meta-modeling method for constructing response surface of outputs (Staum 2009). While the Kriging model is capable of modeling multivariate response, in this paper we use the univariate Kriging model.
We study a kriging-based learning approach to collect information about the AV from on-track experiments. Suppose we are interested in estimating $P(f(\omega)>\gamma)$, where $f:\mathcal X\to\mathbb R$ is an unknown function on $\mathcal X$, $\gamma$ is a given threshold, and $\omega\in\mathcal X$ is a random object under the probability $P$. For instance, $\gamma$ can be $1/b$ and $f(\omega)$ be $\max_{t\in[0,T]}1/R_L(t,\omega)$ in the example described in Section \ref{sec:task}, where $x$ here refers to the set of parameters that controls the human-driving vehicle, which is random and its distribution calibrated from the SPMD database.

To model how information on $f$ updates our estimate on $P(f(\omega)>\gamma)$, we view $f$ as a response surface on the domain $\mathcal X$. We model $f:\mathcal X\to\mathbb R$ as a Gaussian Random Field (GRF) \cite{rasmussen2006gaussian} that is independent of the stochasticity of $\omega$, denoted as\begin{equation*}
	f(\cdot) \sim GRF(\mu(\cdot),\sigma^2(\cdot,\cdot)),
\end{equation*}
where $\mu(\cdot)$ is the mean function and $\sigma^2(\cdot,\cdot)$ is the covariance function of the GRF. Given any fixed design points $x^1,\ldots,x^k\in \mathcal X$, $f(x^1),\ldots,f(x^k)$ comprises a Gaussian random vector with means $\mu(x^i)$ and covariances $\sigma^2(x^i,x^j)$. It is customary to assume that $\mu(x)=b(x)'\beta$ and $\sigma^2(x,\tilde x)=\tau^2 r(x,\tilde x)$ where the correlation function $r(x,\tilde x)$ implies stationary variance over $\mathcal X$ and depends on the design point pairs only through the value of $x-\tilde x$. For simplicity, we will further assume that $\mu(x)=\beta$ for some $\beta \in \mathbb{R}$, which represents a flat belief on $f(\cdot)$ over all the design points. We use the correlation function $r(x,\tilde x)=\exp\{\theta\|x-\tilde x\|^2\} $, where $\|\cdot\|$ denotes the Euclidean norm. This correlation function signifies a higher correlation for test scenarios that are closer to each other. Note that we have adopted intuitive choices for the mean and correlation functions here for convenience, but better ones (in the sense of better reflecting the prior belief on the vehicle behaviors under different test scenarios) should be used with the availability of expert knowledge.

Suppose that the parameters $\beta, \theta, \tau^2$ are known. Given some observations on the value of $f(x)$ at some points in $\mathcal X$, we can update the distribution of $f(\cdot)$ via conditioning. We denote $X$ as the observed design vector $(x^1,...,x^n)$ and $Y$ the associated response vector $(f(x^1),...,f(x^n))$. We define the matrix $\Sigma\in\mathbb R^{n\times n}$ such that its $(i,j)$th entry is $\Sigma_{ij}=\sigma^2(x^i,x^j)$, and define $R=\Sigma/\tau^2$ so that $R_{ij}=r(x^i,x^j)$, for $i=1,...,n$ and $j=1,...,n$. Given observations $(X,Y)$, for any fixed $x\in\mathcal X$, we have \begin{equation*}
	E[f(x)|X,Y]= \beta +r(x|X)'R^{-1}(Y-\beta)
% 	\label{eq:kriging_E}
\end{equation*}
and
\begin{equation*}
	Var(f(x)|X,Y)=\tau^2 (1- r(x|X)'R^{-1}r(x|X)),
% 	\label{eq:kriging_var}
\end{equation*}
where $r(x|X)\in\mathbb R^n$ is a vector with $r(x,x^i)$ as the $i$th element \cite{rasmussen2006gaussian}. Note that $f(x)|X,Y$ still follows a Gaussian distribution. For simplicity, we denote $\mu(x|X,Y)=E(f(x)|X,Y)$ and  $\sigma^2(x|X,Y)=Var(f(x)|X,Y)$. 

In practice, the parameters $\beta, \theta, \tau^2$ need to be either estimated (e.g., by using maximum likelihood) or assigned reasonable values according to expert knowledge. For more details on calibrating the parameters, see, e.g., \shortciteN{ankenman2010stochastic}. In our subsequent discussion, we assume these are given and unchanged throughout the learning process. 
Under the GRF assumption and conditioning on $(X,Y)$, we now set our target quantity of interest as $P(f(\omega)>\gamma|X,Y)$, where $P$ now generates both the stochasticity in $\omega$ and the Gaussian uncertainty in $f$. Typically this probability is larger, i.e., more conservative, than when $f$ is completely known, because of the additional noise coming from the model uncertainty. We view this probability as a reasonable target, but clearly other formulations are plausible.

Note that we have
\begin{equation}
P(f(\omega)>\gamma|X,Y) = E_\omega[P(f(\omega)>\gamma |\omega,X,Y)]
\label{eq:ke_complex}
\end{equation}
where $E_\omega[\cdot]$ denotes the expectation taken with respect to the stochasticity of $\omega$. Since $f(x)|X,Y$ follows a Gaussian distribution with mean $\mu(x|X,Y)$ and variance $\sigma^2(x|X,Y)$, we can write \eqref{eq:ke_complex} further as
% We further expand the integrand as
% \begin{equation}
% E\left[I_{\gamma}(x) \right]=P(y(x)\geq \gamma)=1-\Phi(\frac{\gamma-\mu(x|X,Y)}{\sqrt{\sigma^2(x|X,Y)}}),
% \end{equation}
% Therefore, we write (\ref{eq:ke_complex}) as 
\begin{equation}
E_\omega\left[\bar\Phi\left(\frac{\gamma-\mu(\omega|X,Y)}{\sigma(\omega|X,Y)}\right)\right]
\label{main expression}
\end{equation}
where $\bar\Phi(\cdot)$ denotes the tail distribution function of a standard Gaussian distribution. 
% Now the problem is how to efficiently collect the observations. Since we assume that the experiments are expensive, we want to carefully design the experiments. (see Sections \ref{sec:optim} and \ref{sec:outline} for details.)

\section{SEQUENTIAL SELECTION OF TEST SCENARIOS VIA OPTIMIZATION} \label{sec:optim}

% Here, we assume that a initial observation set $(X_n,Y_n)$ with $n$ samples is given and we know the parameters for the Kriging model. Our goal is to design the next sampling point. The sampling scheme is based on an optimization problem which obtains the sample point with the maximum impact on the event probability estimation using the Kriging model. We measure the impact by the expected ``distance'' between the current model and the model with a new observation. The idea is similar to the knowledge gradient in optimal learning (Frazier et al. 2008).  

From \eqref{main expression}, we design a procedure to sequentially look for the next scenario, or design point, to test the value of $f$ that can in a sense maximize the information gain. We define information gain as the distance between the current estimate of $P(f(\omega)>\gamma|X,Y)$ and its update taking into account the outcome of the next test. We maximize the expected distance under the current posterior distribution. This framework follows generally from the concept of knowledge gradient \cite{powell2012optimal}, but here we are interested in a pure estimation problem instead of an optimization problem. Note that the distribution of $\omega$ is estimated from data, which can be parametrically modeled or fully data-driven, i.e., nonparametric. In general we need to run simulation to evaluate our target quantity, even though $f$ is highly structured.

We present some further notations. Let $(X_n,Y_n)$ be the vectors of historical design points and responses from $f$ collected up to step $n$. We denote $E_n[\cdot]=E[\cdot|X_n,Y_n]$. In particular, $f(x)$ under $E_n[\cdot]$ follows a Gaussian distribution with mean $\mu(x|X_n,Y_n)$ and variance $\sigma^2(x|X_n,Y_n)$. For simplicity, we write $\mu_n(\cdot)=\mu(\cdot|X_n,Y_n)$ and $\sigma^2_n(\cdot)=\sigma^2(\cdot|X_n,Y_n)$.

% $(X_{n+1},Y_{n+1})$ denotes the observation set containing the original sample set $(X_n,Y_n)$ and a new sample $(x,y)$, where $y$ is a realization of random variable $y(x)$. $E_{n+1}[\cdot]$ represents the expectation over $y(x)$, which follows a Gaussian distribution with mean $\mu(x|X_{n+1},Y_{n+1})$ and variance $\sigma^2(x|X_{n+1},Y_{n+1})$. We use $P_n$ to represent the event probability estimation using the Kriging model with the existing sample set $(X_n,Y_n)$. $P_{n+1}$ represents the probability estimation using the Kriging model with the new sample set $(X_{n+1},Y_{n+1})$. 

Let $P_n=P(f(\omega)>\gamma|X_n,Y_n)$ be the current target estimate, and $P_n(x,y)=P(f(\omega)>\gamma|(X_n,x),(Y_n,y))$ be the target estimate if one tests an additional design point $x$ and collects a response $y$. Let $d(\cdot,\cdot)$ be some distance criterion between two probabilities. Given $X_n,Y_n$, we search for the next design point by looking for $x\in\mathcal X$ that solves
% The optimization for searching design point $x$ is
\begin{equation}
	\max_{x \in \mathcal{X}} E_n \left[ d(P_n,P_n(x,f(x)))  \right].
    \label{obj}
\end{equation}
% where $\mathcal{X}$ is the feasible region of $x$ and $d$ is a distance metric. Note that the expectation is over the realization $y$ which follows a Gaussian distribution with mean $\mu(x|X_{n},Y_{n})$ and variance $\sigma^2(x|X_{n},Y_{n})$. Note that we have \begin{equation}
% 	P_n=\int E_n\left[I_{\gamma}(\omega) \right] dF(\omega)=1- \int \Phi(\frac{\gamma-\mu(x|X_{n},Y_{n})}{\sqrt{\sigma^2(x|X_{n},Y_{n})}})  dF(x)
%     \label{obj2_1}
% \end{equation}
% and 
% \begin{equation}
% P_{n+1}=\int E_{n+1}\left[I_{\gamma}(\omega) \right] dF(\omega)=1- \int \Phi(\frac{\gamma-\mu(x|X_{n+1},Y_{n+1})}{\sqrt{\sigma^2(x|X_{n+1},Y_{n+1})}})  dF(x).
% \label{obj2_2}
% \end{equation}

% For simplification, we denote $\mu(\cdot|X_{n},Y_{n})$ as $\mu_n(\cdot)$, $\sigma^2(x|X_{n},Y_{n})$ as $\sigma^2_n(\cdot)$. Similarly, we have $\mu_{n+1}(\cdot)$ and $\sigma^2_{n+1}(\cdot)$ for $\mu(x|X_{n+1},Y_{n+1})$ and $\sigma^2(x|X_{n+1},Y_{n+1})$, respectively. Then we have the objective as
A simple example of $d$ is the squared $L_2$-distance, which we adopt in the sequel. Optimization \eqref{obj} becomes
\begin{align}
&\max_{x \in \mathcal{X}} E_n \left[ (P_n-P_n(x,f(x)))^2  \right]\notag\\
=& \max_{x \in \mathcal{X}} \int (P_n-P_n(x,y))^2 d \Phi\left(\frac{y-\mu_n(x)}{\sigma_n(x)}\right)\notag\\
=& \max_{x \in \mathcal{X}} \int \left(\int \left(\Phi\left(\frac{\gamma-\mu_n(\omega|x,y)}{\sigma_n(\omega|x,y)}\right) -\Phi\left(\frac{\gamma-\mu_{n}(\omega)}{\sigma_{n}(\omega)}\right)\right) dF(\omega)\right)^2 d \Phi\left(\frac{y-\mu_n(x)}{\sigma_n(x)}\right)
\label{eq:obj}
\end{align}
where we denote $\mu_n(\cdot|x,y)=\mu(\cdot|(X_n,x),(Y_n,y))$, $\sigma_n^2(\cdot|x,y)=\sigma^2(\cdot|(X_n,x),(Y_n,y)), $ $F(\cdot)$ the distribution function of $\omega$, and $\Phi(\cdot)$ the standard Gaussian distribution function. 

Note that \eqref{eq:obj} generally does not support closed-form evaluation, and requires running simulation. If $\mathcal X$ is a discrete space, ranking and selection methods can be applied (an approach taken by \shortciteN{Huang2017kriging}). Here we focus on a continuous space for $\mathcal X\subset\mathcal R^d$. We use stochastic approximation (SA) \cite{kushner2003stochastic} to search for a local optimum for \eqref{eq:obj}. This approach follows from \shortciteN{wang2016parallel} that considers parallel Bayesian global optimization where the one-step optimum cannot be solved in closed-form under Gaussian process function models. Note that, like the setting in \shortciteN{wang2016parallel}, since there is no guarantee that the objective function in \eqref{eq:obj} is concave, we can only ensure that our SA converges to a local optimum under suitable conditions.

% The closed form for this objective function is not obvious. One simplified approach is to avoid optimizing the objective function in continuous space. By discretizing the feasible region $\mathcal{X}$, we can estimate the objective function by simulation. Ranking and selection methods can be applied for searching the optimal $x$ on the discretized space.

% We also consider using the stochastic gradient ascent method to search for the optimal value of $x$ on continuous space. We estimate the stochastic gradient using infinitesimal perturbation analysis (IPS). 
We describe our stochastic gradient estimator for the objective function \eqref{eq:obj}. Given i.i.d. samples $\omega_1,...,\omega_m$ drawn from $F$ and $z_1,\ldots,z_m$ drawn from a standard Gaussian distribution, our gradient estimator is a vector in $\mathbb R^d$ whose $j$-th element is given by
\begin{equation}
\frac{1}{m}\sum_{i=1}^{m} -2\left(P_n-P_{n}(x,\sqrt{\sigma_n^2(x)}z_i+\mu_n(x))\right)\frac{\partial}{\partial x_j} \Phi \left( \frac{\gamma-\mu_n(\omega_i|x,\sqrt{\sigma_n^2(x)}z_i+\mu_n(x))}{\sqrt{\sigma^2_n(\omega_i|x,\sqrt{\sigma_n^2(x)}z_i+\mu_n(x))}}\right) \label{grad estimate}
\end{equation}
where $\frac{\partial}{\partial x_j} \Phi \left( \frac{\gamma-\mu_n(\omega|x,y)}{\sqrt{\sigma^2_n(\omega|x,y))}}\right)$ is given by \begin{equation*}
\frac{\partial}{\partial x_j} \Phi \left(\frac{\gamma-\mu_n(\omega|x,y)}{\sqrt{\sigma^2_n(\omega|x,y)}}\right)=-\phi \left( \frac{\gamma-\mu_n(\omega|x,y)}{\sqrt{\sigma^2_n(\omega|x,y)}}\right)\left( \frac{1}{2} \sigma^2_n(\omega|x,y) ^{-3/2} \frac{\partial}{\partial x_j} \sigma^2_n(\omega|x,y)+\sigma^2_n(\omega|x,y) ^{-1/2} \frac{\partial}{\partial x_j} \mu_n(\omega|x,y)\right).
% \label{eq:derivative}
\end{equation*}
Here we have \begin{equation*}
\frac{\partial}{\partial x_j} \mu_n(\omega|x,y)=\frac{\partial}{\partial x_j} r_n(\omega|x) R_{n}(x)^{-1} [(Y_{n},y)-\beta]+ r_n(\omega|x) \frac{\partial}{\partial x_j}R_{n}(x)^{-1} [(Y_{n},y)-\beta]+ r_n(\omega|x) R_{n}(x)^{-1} \frac{\partial}{\partial x_j}(Y_{n},y)
\end{equation*} and \begin{equation*}
 \frac{\partial}{\partial x_j} \sigma^2_n(\omega|x,y) =-\tau^2 \left(2 \frac{\partial}{\partial x_j} r_n(\omega|x)'R_{n}(x)^{-1} r_n(\omega|x) +r_n(\omega|x)'\frac{\partial}{\partial x_j} R_{n}(x)^{-1} r_n(\omega|x) \right),
\end{equation*}
where we use $r_n(\omega|x)=r(\omega|(X_n,x))\in\mathbb R^{n+1}$ to denote the vector whose $i$th element is $r(\omega,x^i)$ for $i=1,\ldots,n$ and $(n+1)$th element is $r(\omega,x)$, $R_{n}(x)\in\mathbb R^{(n+1)\times(n+1)}$ to denote the matrix whose $(i,j)$th entry is $r(x^i,x^j)$ for $i=1,...,n$ and $j=1,...,n$, $(i,n+1)$th entry is $r(x^i,x)$ for $i=1,...,n$, $(n+1,j)$th entry is $r(x,x^j)$ for $j=1,...,n$, and $(n+1,n+1)$th entry is $r(x,x)$.

Furthermore, we have \begin{equation*}
\frac{\partial}{\partial x_j} R_{n}(x)^{-1}= R_{n}(x)^{-1} \frac{\partial}{\partial x_j} R_{n}(x) R_{n}(x)^{-1}.
\end{equation*}
The vector $\frac{\partial}{\partial x_j} r_n(\omega|x)$ has 0 in all entries but the last, which is equal to $\frac{\partial}{\partial x_j} r(\omega,x)$. $\frac{\partial}{\partial x_j} R_{n}(x)$ has 0 in all entries except the last row and column, where the $(i,n+1)$th entry and $(n+1,i)$th entry is equal to $\frac{\partial}{\partial x_j} r(x^i,x)$ for $i=1,...,n$ where $x^i$ denotes the $i$th observation. The vector $\frac{\partial}{\partial x_j}(Y_{n},y)$ has 0 in all entries but the last, which is equal to \begin{equation*}
\frac{1}{2}  \sigma^2_{n}(x) ^{-1/2} z \frac{\partial}{\partial x_j}\sigma^2_{n}(x)+\frac{\partial}{\partial x_j} \mu_{n}(x).
\end{equation*}
Lastly, we have \begin{equation*}
\frac{\partial}{\partial x_j} \mu_{n}(x)= \frac{\partial}{\partial x_j} r_{n} (x) R_{n}^{-1} [Y_{n}-\beta]
\end{equation*} and \begin{equation*}
\frac{\partial}{\partial x_j}\sigma^2_{n}(x)=-\tau^2 \left(2 \frac{\partial}{\partial x_j} r_{n}(x)'R_{n}^{-1} r_{n} (x) \right),
\end{equation*}
where \begin{equation*}
\frac{\partial}{\partial x_j} r_{n}(x)=\left[  {\begin{array}{c}
   \frac{\partial}{\partial x_j} r(x^1,x)  \\
   \frac{\partial}{\partial x_j} r(x^2,x) \\
   ...\\
   \frac{\partial}{\partial x_j} r(x^n,x)
  \end{array} }  \right]
\end{equation*}
and $R_n\in\mathbb R^{n\times n}$ is a matrix whose $(i,j)$th entry is $r(x^i,x^j)$ for $i=1,...,n$ and $j=1,...,n$, $r_n(x)\in\mathbb R^n$ is a vector whose $i$th element is $r(x^i,x)$ for $i=1,...,n$, and \begin{equation*}
\frac{\partial}{\partial x_j} r(x^i,x)=r(x^i,x)(-2 \theta(x^i_j-x_j)).
\end{equation*}
The above estimator is only a heuristic that roughly resembles an infinitesimal perturbation analysis. Upon closer inspection, one can see that the term $\sigma_n^2(\omega|x,y)$ in the denominator in the formulas above is close to 0 if $\omega$ approaches any of the observed design points, a consequence of the fact that the responses at those points are completely known. This may blow up the gradient estimate. This issue can potentially be addressed by adding artificial small noise to the kriging model to inflate the variance from zero at those positions. An alternative is to use the finite-difference method, although this will reduce the efficiency of the resulting gradient descent algorithm.

With the gradient estimator, we iterate\begin{equation}
x^{(k+1)}=x^{(k)}+a_kg^{(k)}(x^{(k)})
 \label{eq:sa}
\end{equation}
where $g^{(k)}(x)$ denotes the gradient estimator in \eqref{grad estimate}, for $k=1,2,\ldots$ starting from an initial solution $x^{(0)}$ , to optimize the objective function in \eqref{eq:obj} according to a heuristic Robbins-Monro SA. The step size is taken as $a_k=a_0/k$. One may also apply the algorithm at multiple starting points in view of the non-convexity of the problem. 
% One can refer to \citeN{asmussen2007stochastic} Chapter VIII for the behaviors and features of SA using simulation-based gradient estimator. 

% update \cite{robbins1951stochastic}.
% \section{Outline of the Adaptive Sampling Scheme} \label{sec:outline}
% Note that the optimization objective function in Section \ref{sec:optim} is based on an initial observation set. To apply the optimal scheme, we need to design the initial experiments. Therefore, we use standard Kriging design method for the initial experiments. For example, we can use the Latin Hypercube Sampling method. Once we have the initial sample set, we can sequentially select the optimal sample points using the optimization function. 
Overall, to sequentially select the design points, the steps consist of:
\begin{enumerate}
	\item Use a small-sample space-filling design to build an initial observation set $(X_0,Y_0)$ and construct an initial kriging model. 
    \item Approximately solve (\ref{obj}) to select the next design point $x^*$. This involves recursion using \eqref{eq:sa} where $g^{(k)}(x^{(k)}),k=1,2,\ldots$ are estimated by generating i.i.d. samples $\omega_1,\ldots,\omega_m$ from $F$ and $z_1,\ldots,z_m$ from standard Gaussian as described above.  \label{item:step2}
    \item Conduct an experiment at $x^*$ and add $x^*$ and the associated experimental outcome to the observation set $(X_n,Y_n)$ to get $(X_{n+1},Y_{n+1})$.\label{item:step3} 
        \item Update the kriging model using the observation set $(X_{n+1},Y_{n+1})$. \label{item:step4}
	    \item Repeat steps \ref{item:step2}, \ref{item:step3} and \ref{item:step4} until the kriging model is acceptable.
\end{enumerate}

% We note that we can use the discretization approach or the stochastic gradient approach for step \ref{item:step3}. Using the discretization approach requires a decent method to discretize the sampling space and an appropriate ranking and selection scheme to find the optimal point. The stochastic gradient approach searches for the optimal sampling point on continuous space, while the algorithm cannot guarantee the convergence to optimal. The stochastic feature and the selection of step parameter introduce complexity to the implementation. 

% The algorithm might converge to a local optimum or to a non-optimal point. To tackle these problems, we need to carefully set up the algorithm. We can use multiple start points for the search algorithm and compare the obtained points. We evaluate the performance of the stochastic gradient approach in Section \ref{sec:numerical}.

\section{NUMERICAL EXAMPLES} \label{sec:numerical}
This section shows some numerics on our information criterion and simple illustrations of our procedure.

\subsection{Illustration of the Information Criterion}

% Once we have a Kriging model, we can use it to predict the outcome at any given point. Adding one more sample to the current Kriging model will change the prediction of the Kriging model and therefore change the probability estimation using the Kriging model. We defined the information gain in Section \ref{sec:optim} to search for design points. In this section, we want illustrate the interpretation of the information gain.

Here we present an example to illustrate the intuition behind the proposed information criterion. By contrasting with a simple alternative criterion, we demonstrate the relation between the proposed criterion and the underlying probability distribution.

Consider the generic target probability of interest $P(f(\omega)>\gamma)$, where $\omega \in \mathcal{X}$ is a random object with probability $P$. In addition to \eqref{obj}, we consider an alternative criterion to select the next design point by maximizing the pointwise variance of $I(f(x)>\gamma)$ over $x\in \mathcal{X}$ under the posterior distribution on $f$, namely $f(x)\sim N(\mu_n(x),\sigma_n^2(x)$), where $I(\cdot)$ is the indicator function. In other words, we maximize 
\begin{equation}
E_n\left( I\left(f(x) > \gamma\right)- P_n\left(f(x) > \gamma\right) \right)^2\label{new criterion}
\end{equation}
where $E_n[\cdot]$ and $P_n(\cdot)$ refer to the conditional distribution on $X_n,Y_n$ as before. Criterion \eqref{new criterion}, which we name the local prediction impact for convenience, does not depend on the distribution of $\omega$ but only measures the uncertainty (or confidence of our knowledge) on the values of the function $f(\cdot)$ at different points. This contrasts our suggestion in \eqref{obj} that accounts for both the uncertainty on $f$ and the distribution of $\omega$, and in this sense \eqref{new criterion} is a less comprehensive measure. In general, one would expect that the information gain measured by \eqref{obj} is large when the local prediction impact is large and the position of interest is ``important" according to the distribution of $\omega$. On the other hand, a position with a large local prediction impact may not necessarily be important in determining the estimate of $P(f(\omega)>\gamma)$, since the latter depends on the distribution of $\omega$. 

% lead to a gain On the other hand, a point with large local impact can have small information gain if the underlying density of the point is rare. Here, we use a lane change scenario example to show this interpretation.

% We interpret the information gain from two aspects: the local prediction impact and the underlying distribution of the design points. When
% First, we propose an intuitive measure for the impact on local prediction. We want to use the Kriging model to interpret the difference between the prediction and possible outcome at a given point. We consider that if we sample at an arbitrary point $x$ and we are interested in the outcome $I(f(x) > \gamma)$. Note that $f(x)|(X,Y)$ is random regarding the Kriging model, we can interpret that the outcome $I(f(x) > \gamma)$ of a sample at $x$   is a random variable. We want to measure the difference between this outcome and the prediction of the Kriging prediction $ P\left[f(x) > \gamma|(X,Y) \right]$ on $x$. We take the squared of the difference and integral over all possible $f(x)$. This gives us \begin{equation}
% E\left[ \left( I\left(f(x) > \gamma|(X,Y) \right)- E\left[f(x) > \gamma|(X,Y) \right] \right)^2 |(X,Y)\right]
% \end{equation}
% Note that this is the variance of $I\left(f(x) > \gamma|(X,Y) \right)$. Considering that a larger variance of $I\left(f(x) > \gamma|(X,Y) \right)$ indicates that the Kriging gives less assured prediction on $x$, it is reasonable to use the variance to measure the impact on a local point.

\begin{figure}[ht]
							\centering
							
							\begin{minipage}[b]{0.4\textwidth}
								\centering
								\includegraphics[width=\linewidth]{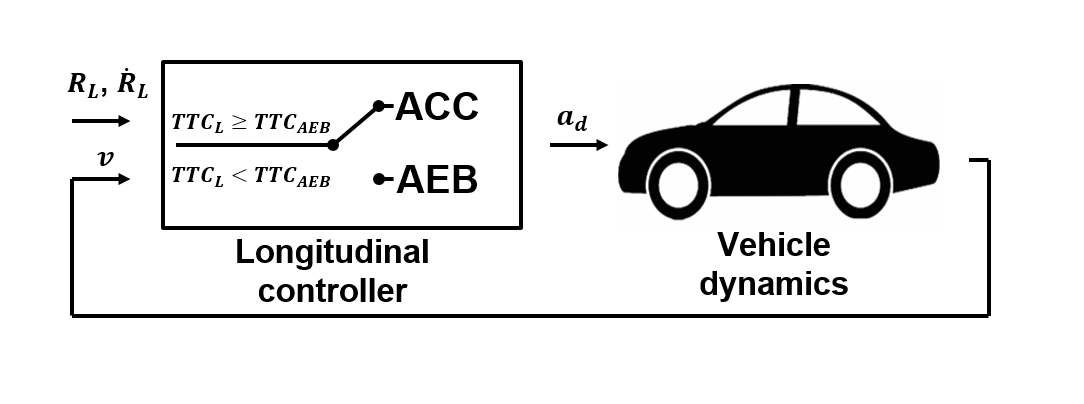}
								\caption{An example of AV control mechanism.}
								\label{fig:av_model}
							\end{minipage}
                            \begin{minipage}[b]{0.4\textwidth}
								\centering
                            \includegraphics[width=\linewidth]{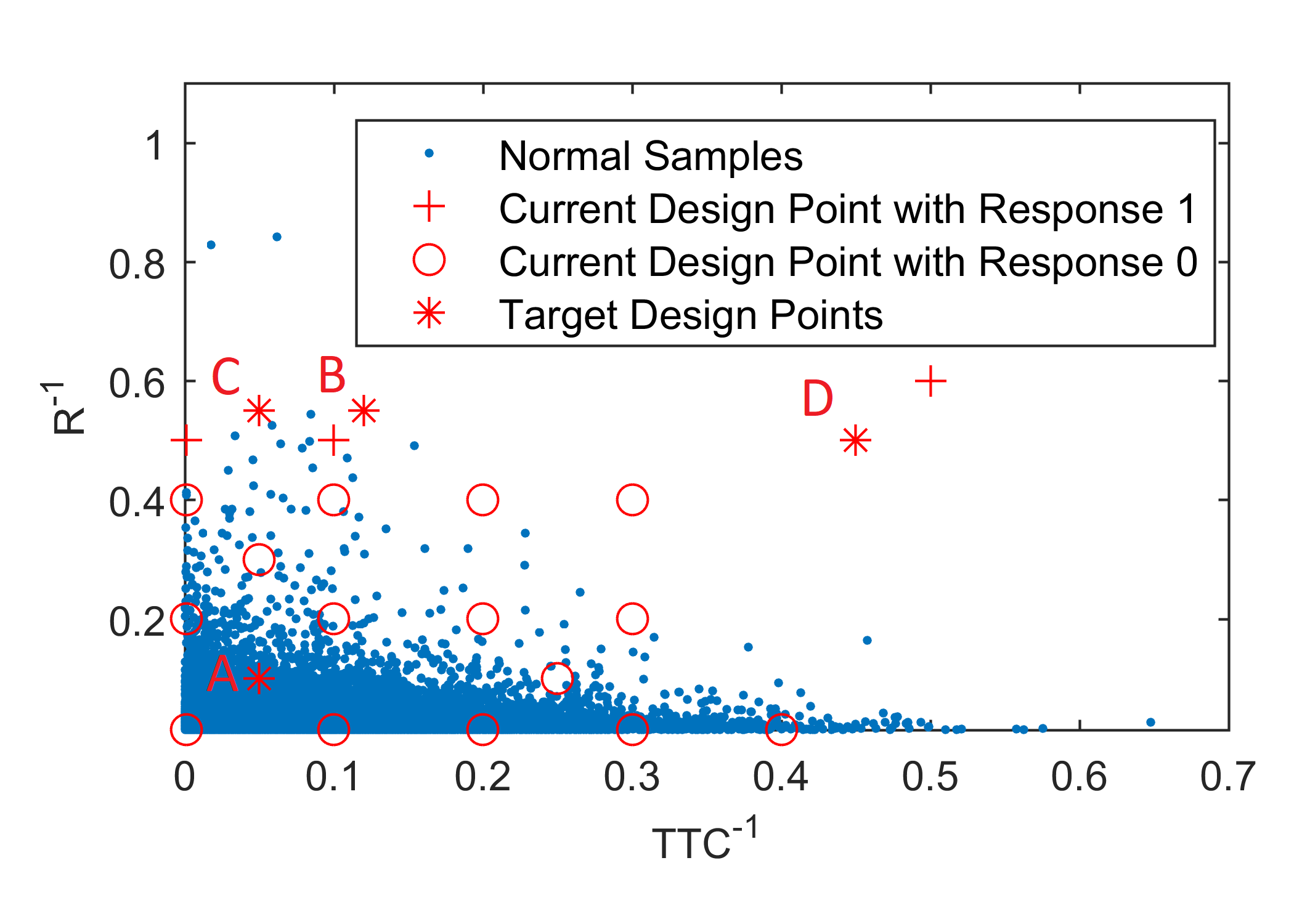}
	\caption{Prediction of a kriging model in the lane change setting.}
	\label{fig:exp3_set}
							\end{minipage}%
			\end{figure}

We demonstrate the two criteria with a study of lane change scenarios described in Section \ref{sec:task}. We assume that the AV uses a deterministic system with Adaptive Cruise Control (ACC) and Autonomous Emergency Braking (AEB) \shortcite{ulsoy2012automotive} (see Fig. \ref{fig:av_model}), but this is supposedly unknown to the tester. 
% With a random generated scenario, we can simulate the interaction of the vehicles in the lane change procedure and determine whether a crash will happen. This model allows us to use Monte Carlo simulation to estimate the probability of certain events (generally risk events) occcurring during the lane change scenario. Previously \shortciteN{zhao2017accelerated} and \shortciteN{huang2017accelerated}, we evaluated the probability of crash, conflict, and injury as the events of interest.
There are three key variables that constitute the scenario, namely the frontal vehicle's velocity $v$, range $R$ and time to collision $TTC$, where we define $TTC$ as
\begin{equation*}
	TTC=- \frac{R}{\dot{R}}.
\end{equation*}
As described in \shortciteN{zhao2017accelerated}, when the velocity $v$ is between 5 to 15 $m/s$, the other two variables $R^{-1}$ and $TTC^{-1}$ are independent of each other.  $TTC^{-1}$ can be modeled by an exponential distribution and $R^{-1}$ by a Pareto distribution. Here we define $\omega=[TTC^{-1},R^{-1}]$, and we are interested in estimating $P(\max_{t\in[0,T]}1/R_L(t)>1/2)$, i.e., the probability that the two vehicles have a minimum range smaller than 2 meters, when the velocity of the leading vehicle $v$ is set to lie in the aforementioned range. 

We use a kriging model with parameters $\beta=0$, $\tau^2=0.01$ and $\theta=50$. We set the prediction threshold $\gamma=0.5$ for simplicity. The zero mean of $\beta$ is chosen to reflect the belief that the response of a scenario with no information is far from being a critical event. We choose the value of $\tau^2$ which intuitively puts $\gamma=0.5$ to be three standard deviations higher than the mean of a scenario $x$ that has no information (i.e., $f(x) \sim N(0,0.01)$). $\theta=50$ is selected to make the correlation between scenarios with distance $0.05$ (believed to represent initial conditions with different AV behaviors) to be small enough (less than $0.01$). 

We use 20 initial design points to build the model and its value of $I(f(x)>\gamma)$ is shown in Fig. \ref{fig:exp3_set}. The blue dots represent a sample distribution of the variable $\omega$. Red circles are existing design points with return 0 and red crosses are existing design points with return 1. We consider four arbitrarily picked new design points (which we call points A, B, C and D) shown by the red stars, whose coordinates are shown in the first row of Table \ref{table:exp3}. The local prediction impacts and the information gains depicted by the objective in \eqref{obj} of these design points are shown in the second and third rows respectively.
% The four target design points are represent by the red stars. 

We see that points B and C have smaller local prediction impacts than points A and D, which can be attributed to the vicinity of their positions to those of the historical data that subsequently reduces the uncertainty of $f$. This translates to a smaller variance of $I(f(x)>\gamma)$ and hence a smaller local prediction impacts. Relatedly, these points also have a low information gain measured by \eqref{obj}. However, point D, even though far away from the positions of the historical data, has an even lower information gain. This can be attributed to the tiny density of $\omega$ at this point, which makes the overall information gain low. In contrary, point A has a higher density of $\omega$ and consequently a higher information gain.

% the correlation is large between these points and the known data points. The variance of the predictions is low which indicates that these two points have relatively low impact on local prediction. We observe that both points A and D have large variance on $I(f(x)>\gamma)$, which means that sampling at this point has a high impact on local prediction.

% The information gains for points B and C are low. We interpret that the low information gain is caused by both low distribution density and local prediction impact. While points A and D have large impact on local prediction, since the distribution density for point D is too rare, the information gain is even smaller than points B and C. While point A has a large distribution density, so the information gain is larger than other points. 

\begin{table}[ht]
\centering
\caption{Local prediction impacts and information gains of 4 arbitrarily picked design points.}
\renewcommand{\arraystretch}{1.5}
\label{table:exp3}
%\resizebox{0.6\columnwidth}{!}{%
\begin{tabular}{|l|l|l|l|l|}
\hline
        & A          & B           & C           & D          \\ \hline
Coordinate   & (0.05,0.1) & (0.12,0.55) & (0.05,0.55) & (0.45,0.5) \\ \hline
Local prediction impact & 0.209   & 0.0996    & 0.0471   & 0.249   \\ \hline
Information gain    & 0.0082   & $4.26 \times 10^{-8}$    & $6.76 \times 10^{-8}$    & $3.13 \times 10^{-11}$   \\ \hline
\end{tabular}%}
\end{table}

\subsection{Example of the Sequential Learning Approach}
To illustrate our sequential learning approach, we use a simple hypothetical problem where we define the probability of interest as $P(\omega_1 + \omega_2 > 2)$, with two random objects $\omega_1,\omega_2$ each following a standard Gaussian distribution (in the lane change scenario described before $\omega_1,\omega_2$ would correspond to the initial conditions such as frontal vehicle velocity, with a correspondingly more sophisticated $f$ function). The true probability is $1- \Phi (\frac{2}{\sqrt{2}}) \approx 0.0786$.

We use a kriging model with parameters $\beta=0$, $\tau^2=1$, and $\theta=1$. Here the parameters are arbitrarily chosen, as we assume that no prior information is available. We start with 20 initial design points. In the SA scheme, we use $a_k=a_0/k$ as the step size parameter with $a_0=20$, and we terminate the scheme after 50 iterations, at each new design point. The gradient estimator is averaged from 1,000 samples. For illustration, we use 10,000 samples to estimate the target probability under the kriging model to assess its error relative to the truth.  

\begin{figure}[ht]
							\centering
							\begin{minipage}[b]{0.5\textwidth}
								\centering
                                \captionsetup{width=.9\linewidth}
	\includegraphics[width=\linewidth]{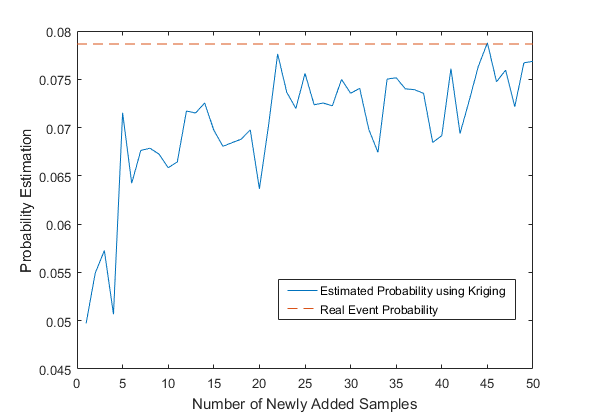}
	\caption{Changes in probability estimates as new observations are collected to update the kriging model.}
	\label{fig:est_prob}
							\end{minipage}%
							\begin{minipage}[b]{0.5\textwidth}
								\centering
                                \captionsetup{width=.9\linewidth}
	\includegraphics[width=\linewidth]{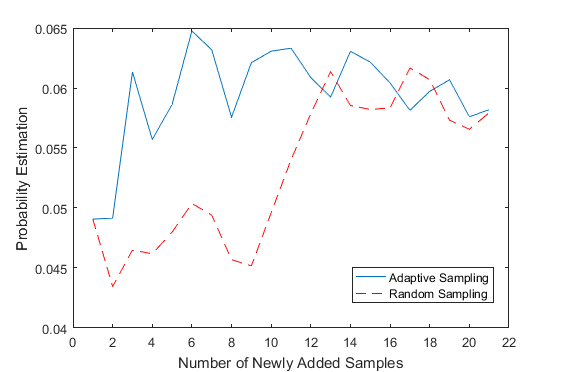}
	\caption{Comparison between probability estimates of the learning approach and random sampling without optimization.}
	\label{fig:compare}
							\end{minipage}
			\end{figure}

\begin{figure}[ht]
							\centering
							\begin{minipage}[b]{0.5\textwidth}
								\centering
                                \captionsetup{width=.9\linewidth}
	\includegraphics[width=\linewidth]{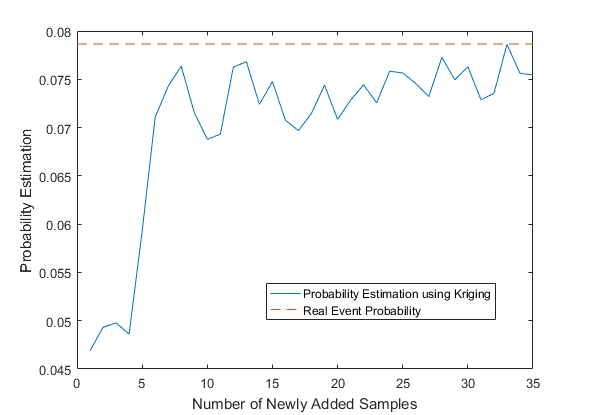}
	\caption{Sequential selection of design points using SA with starting point fixed at $(1,1)$.}
	\label{fig:stationary1}
							\end{minipage}%
							\begin{minipage}[b]{0.5\textwidth}
								\centering
                                \captionsetup{width=.9\linewidth}
	\includegraphics[width=\linewidth]{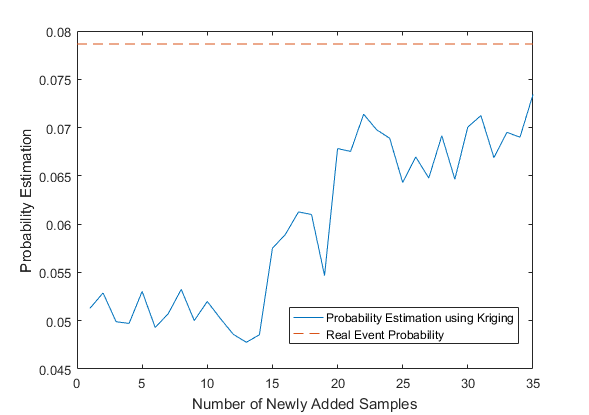}
	\caption{Sequential selection of design points using SA with starting point fixed at $(0,0)$.}
	\label{fig:stationary2}
							\end{minipage}
			\end{figure}

Fig. \ref{fig:est_prob} shows that as we collect more observations to update the kriging model, the probability estimate gradually converges to the true probability. At each step, we start the SA from a randomly generated point using a standard Gaussian distribution. To illustrate the benefit from the optimization step, Fig. \ref{fig:compare} compares our approach with random sampling at each step, where this random sample is precisely the starting point of our SA scheme. We observe that our sequential learning approach converges to the true probability quickly in the first few steps, but the convergence slows down as the learning progresses. This may be caused by a saturation in terms of the highest accuracy affordable by the SA's noises, as well as the heuristic nature of our approach. Finally, Fig. \ref{fig:stationary1} and \ref{fig:stationary2} show the probability estimates when we use SA with starting points fixed at $(1,1)$ and $(0,0)$ respectively, at each learning step. We see that the probability estimates move towards the truth regardless of the starting points, giving a sign that the SA algorithm is at least working. Moreover, starting from $(1,1)$ appears to achieve faster convergence, which can be reasoned by the fact that $(1,1)$ is closer to the boundary of the event $\omega_1+\omega_2>2$ that facilitates the involved learning process. We note that the lines in the figures appear a bit fluctuant as they are illustrated in the scale of the probability estimates, which is small relative to the simulation replication size we use to generate them (i.e., $10,000$). Further investigation is clearly needed, but the above observations aim to show some preliminary insights on the behavior of our approach and confirm its potential.

% To further validate the proposed sampling approach, we fix the start point for the stochastic approximation in each iteration. Fig.We observe that $(1,1)$ is a better start point, since the convergence of estimation takes less steps then starting from $(0,0)$. This might because starting from $(1,1)$ gives better local optimum. The proposed approach is valid for both start points.

\section{CONCLUSION} \label{sec:conclusion}

This paper presents a sequential learning approach based on using kriging models to approximate AV behaviors, to reduce on-track experimentation for AV safety evaluation. The approach relies
on a heuristic simulation-based gradient descent procedure to search for the best next test scenario in terms of maximizing an information criterion regarding the accuracy of conflict probability evaluation. We derive a gradient estimator and investigate the performance of our procedure. Numerical examples show that our approach sequentially improves our probability estimate, and appears to perform better than simple strategies such as random scenario sampling. Future work includes the studies of further assumptions of the kriging models in the AV evaluation context and developments of scenario search procedures that are both more efficient and theoretically sound.

% However, the fluctuancy of the prediction indicates that the method is instable in selecting the ``optimal'' test scenario. This might be caused by the stochasticity or the blow-up issue of the gradient estimator.

\section*{ACKNOWLEDGMENTS}
We gratefully acknowledge support from the National Science Foundation under grants CMMI-1542020, CMMI-1523453 and CAREER CMMI-1653339.

% The authors acknowledge support from the University of Michigan Mobility Transformation Center under grant number N021552.

% Please don't exchange the bibliographystyle style
% \bibliographystyle{wsc}
% % AUTHOR: Include your bib file here
% \bibliography{Mendeley_MTC_AE}

\bibliographystyle{wsc}
 \bibliography{wsc_citation.bib}

\section*{AUTHOR BIOGRAPHIES}
\noindent {\bf ZHIYUAN HUANG} is a second-year Ph.D. student in the Department of Industrial and Operations Engineering at the University of Michigan, Ann Arbor. His research interests include simulation and stochastic optimization. His email address is \email{zhyhuang@umich.edu}.\\

\noindent {\bf HENRY LAM} is an Assistant Professor in the Department of Industrial and Operations Engineering at the University of Michigan, Ann Arbor. His research focuses on stochastic simulation, risk analysis, and
simulation optimization. His email address is \email{khlam@umich.edu}.\\

\noindent {\bf Ding Zhao} is a Assistant Research Scientist in the Department of Mechanical Engineering at the University of Michigan, Ann Arbor. His research focuses on Connected and Automated Vehicles (CAVs) using synthesized approaches rooted in advanced statistics, modeling, optimization, dynamic control, and big data analysis. His email address is \email{zhaoding@umich.edu}.\\

\end{document}